\documentstyle[preprint,aps]{revtex}

\begin{document}
\draft

\title{The 1/3-shot noise suppression in diffusive nanowires}
\author{M.~Henny, S.~Oberholzer, C.~Strunk and C.~Sch\"{o}nenberger}
\address{Institut f\"{u}r Physik, Universit\"{a}t Basel, Klingelbergstr. 82,\\
CH-4056 Basel, Switzerland}
\date{\today}
\maketitle

\begin{abstract}
We report low-temperature shot noise measurements of short diffusive Au
wires attached to electron reservoirs of varying sizes. The measured noise
suppression factor compared to the classical noise value $2e\left| I\right| $
strongly depends on the electric heat conductance of the reservoirs. For
small reservoirs injection of hot electrons increases the measured noise and
hence the suppression factor. The universal $1/3$-suppression factor can
only asymptotically be reached for macroscopically large and thick electron
reservoirs. A heating model based on the Wiedemann-Franz law is used to
explain this effect.
\end{abstract}

\vspace{.5cm}
\vspace{.5cm}

\section{Introduction}

The granularity of charge flow, due to the discreteness of electrical charge
in units of $e$, causes the electrical current to fluctuate around its
average value $I$. The spectral density of these fluctuations $S_{I}$ are
known as noise.\cite{deJong} In equilibrium ($I=0$) at temperature $T$,
thermal fluctuations give rise to Johnson-Nyquist noise\cite{Johnson} $%
S_{I}=4kT/R$ for a wire with resistance $R$. For a non-equilibrium
situation, in which a net-current flows, excess noise appears in addition to
equilibrium noise. This so called shot noise is directly related to the
degree of randomness in carrier transfer caused by the electron scattering
in the wire. From shot noise one can therefore obtain information on the
conduction mechanism not accessible from conventional resistance
measurements. If the number of transferred electrons in a given time
interval is determined by a Poissonian distribution, the current shows shot
noise with a value given by the well known Schottky formula\cite{Schottky} 
$S_{{\rm Poisson}}=S_{I}=2e\left| I\right| $. This classical shot noise is
observed in tunnel junctions or vacuum tubes, for example.\cite{Birk} Shot
noise for wires connected to electron reservoirs on each end is lower than
the classical shot-noise value $S_{{\rm Poisson}}$ by a factor that depends
on the ratio of the wire length $L$ with respect to characteristic
scattering lengths like the elastic ($l_{e}$), electron-electron ($l_{e-e}$)
and electron-phonon ($l_{e-ph}$) scattering lengths. In a ballistic wire 
($L\ll l_{e}$), shot noise vanishes, since scattering is completely 
absent.\cite{ReznikovKumar} In the diffusive regime ($L\gg l_{e}$), excess 
noise varies linearly with current only if $L\ll l_{e-ph}$. 
Two limiting cases can then be distinguished. 
In the interacting (electron-) regime, i.e.
$L\gg l_{e-e}$, the electrons assume a Fermi-Dirac
distribution with a locally varying temperature above the phonon
temperature. The noise is given by the Johnson-Nyquist noise of the mean
electron temperature averaged over the whole wire length. Independent of
material and geometric parameters, shot noise is reduced by a factor of 
$\sqrt{3}/4$ from the classical value.\cite{Steinbach} 
On the other hand in the non-interacting (electron-) regime,
i.e.  $L\ll  l_{e-e}$, the distribution
function $f$ is no longer a Fermi-Dirac function. For this regime various
theories predict a fundamental shot noise reduction factor of $1/3$. Using
random-matrix theory, Beenakker and B\"{u}ttiker have calculated this factor
first.\cite{Beenakker} In their derivation, the conductor is implicitly
assumed to be phase coherent and the factor is obtained as the
ensemble-averaged value. In a semiclassical picture, where no
phase-coherence is required, the fluctuations of the distribution function 
$f$ yield surprisingly the very same suppression factor.\cite{Nagaev1}
Moreover, the sequential transfer of electrons through a series of tunnel
barriers has also been shown to lead to exactly the same noise reduction
factor of $1/3$ in the limit of a large number of barriers.\cite{deJong2}
The remarkable fact, that the same reduction factor of $1/3$ is derived from
different theoretical models, has been ascribed to a numerical coincidence
by Landauer.\cite{Landauer} Recently, two publications have provided
additional theoretical support for the universality of the magic $1/3$-suppression
factor. In the first, a non-degenerate diffusive conductor \cite{Gonzalez}
is studied by computer simulations, while in the second, the universality is
extended to multiterminal diffusive conductors with arbitrary shape and
dimension.\cite{Eugene}

Despite the remarkable universality of the reduction factor $1/3$ obtained
from various theoretical models for the non-interacting electron-regime, 
a clear experimental
confirmation in the asymptotic
limit $eV\gg kT$, in which shot noise is much larger than thermal noise, is
lacking. To clearly distinguish the non-interacting from the interacting
regime by noise measurements, a relatively high accuracy is needed
allowing to separate the two close-lying reduction factors $1/3$ and $\sqrt{3}/4$
from a measurement of noise, which by itself is a small
quantity of order $10$~pV/$\sqrt{\rm Hz}$.

The first experiment in this field was done by Liefrink {\it et al.}\cite
{Liefrink} using a two-dimensional electron gas, which was electrostatically
confined into a wire. A linear variation of the noise with current was
found. The measured reduction factors however were ranging from $0.2$ to 
$0.45$. Steinbach {\it et al.}\cite{Steinbach} found excellent agreement with
the $\sqrt{3}/4$-theory for a Ag wire of $30$~$\mu $m length, but reported a
value between $1/3$ and $\sqrt{3}/4$ for a $1$~$\mu $m long wire, although
theory\cite{Altshuler} predicts $L\ll l_{e-e}$ for this length at $50$~mK.
Schoelkopf {\it et al.} were the first to study high-frequency (quantum-)
shot noise of diffusive wires.\cite{Schoelkopf} By comparing measured
differential-noise $dS_I/dI$ with the $1/3$ and $\sqrt{3}/4$ theories, good 
agreement was found for the non-interacting regime. 
However, the absolute slope, i.e. the $1/3$-reduction factor, was not 
measured in the asymptotic limit $eV\gg kT$. 
A novel approach enabling to distinguish between the interacting and
the non-interacting regimes, was introduced by H. Pothier {\it et al.}, who
measured directly the distribution function $f(E,x)$ of a wire by tunneling
spectroscopy.\cite{Pothier}

We will show in the present paper that the electron reservoirs connected to the 
wire are of great importance for the confirmation of the $1/3$-suppression 
factor. 
Bounded by the limiting values $1/3$ and $\sqrt{3}/4$,
the measured noise-reduction factor can in principle distinguish between the
non-interacting ($L\ll l_{e-e}$) and the interacting regime ($L\gg l_{e-e}$
). This is, however, only true, if heating in the electron reservoirs is
absent. Our experiments demonstrate, that noise-reduction factors close to 
$\sqrt{3}/4$ can be measured, even though the wires are in the
non-interacting regime! This is demonstrated to be caused by unavoidable
reservoir heating, which results in a significantly increased measured slope 
of the shot noise in the asymptotic limit. 
We discuss noise measurements of three Au
wires that mainly differ in the size of the attached electron reservoirs.
The sample with the thickest reservoirs, i.e. the highest reservoir heat
conductivity, closely approaches the universal $1/3$ shot-noise reduction
factor.

\section{Theory}

\subsection{Noise in diffusive conductors}

The current flowing through a wire exhibits fluctuations $\Delta I=I(t)-I$
around the average current $I$. The spectral density of these
current fluctuations, i.e. current noise, can be written as the Fourier
transform of the current autocorrelation function:\cite{Kogan}

\begin{equation}
S_{I}\left( \omega \right) =2\int_{-\infty }^{\infty }dte^{i\omega t}\langle
\Delta I(t+t_{0})\Delta I(t_{0})\rangle _{t_{0}}  \label{fourauto}
\end{equation}
In thermodynamical equilibrium Eq.~(\ref{fourauto}) yields $S_{I}=4kT/R$,
called thermal or Johnson-Nyquist noise.\cite{Johnson} Under current bias
the individual charge pulses of the electrons give rise to
out-of-equilibrium noise known as shot noise. If the electrons pass rarely
and completely random in time governed by a Poissonian process, one obtains
the classical shot noise $S_{I}=2e\left| I\right| $ as derived by 
Schottky.\cite{Schottky} 
If in contrast the electron stream is denser, correlations
due to many-particle statistics induced by the Pauli principle\cite{Liu} or
due to Coulomb interaction\cite{Birk} can significantly reduce shot noise.
For $\hbar \omega \ll kT$ thermal and shot noise display a white spectrum
(frequency independent). In contrast, resistance fluctuations related to the
dynamics of impurities in the sample display in general so called 
$1/f$-noise proportional to $1/\omega $ over a large frequency 
range.\cite{Kogan}
We restrict ourselves to a frequency range, which is high enough to safely
neglect the $1/f$-noise.

An elegant framework to describe the shot noise power of a mesoscopic device
is the Landauer-B\"{u}ttiker formalism. It is valid in linear response and
in the absence of inelastic scattering. The current is carried by
independent parallel channels with a transmission probability $T_{n}$. The
conductance is then written as $G=\frac{e^{2}}{h}\sum_{n}T_{n}$ and the
shot noise at zero temperature reads:

\begin{equation}
S_{I}=2e\left| V\right| \frac{e^{2}}{h}\sum_{n}T_{n}(1-T_{n})
\label{LaBue}
\end{equation}
A diffusive wire is described as an ensemble of many parallel channels.
Random matrix theory predicts a bimodal distribution function for
transmission probabilities, which leads to a suppression of shot noise by a
factor of $1/3$ compared to its classical value:\cite{Beenakker}

\begin{equation}
S_{I}=\frac{1}{3}2e\left| I\right|  \label{onethird}
\end{equation}

Nagaev proposed a semiclassical approach to determine the noise in a
diffusive wire.\cite{Nagaev1} Starting from a kinetic equation
for the electron occupation probability $f(E,x)$, current noise is shown to
be related to the fluctuations of the occupation number given by $f(1-f)$.
Explicitly, the following equation was derived:

\begin{equation}
S_{I}=4G\left\langle \int_{-\infty }^{\infty }f\left( E,x\right) \left[
1-f\left( E,x\right) \right] dE\right\rangle _{wire}  \label{Nagaev}
\end{equation}
In this approach, phase coherence is not required in contrast to random
matrix theory. Furthermore, it has the advantage that inelastic scattering
processes can easily be included. 
%The occupation probability $f(E,x)$ is obtained by solving a differential 
%equation for $f$. 
They are introduced by scattering integrals $I_{ee}$ for
electron-electron scattering and $I_{ph}$ for electron-phonon scattering.
%Both $I_{ee}$ and $I_{ph}$ are calculated by integrating over all states
%contributing to the scattering process.
$f$ can be obtained by the following diffusion equation:

\begin{equation}
D\frac{d^{2}}{dx^{2}}f\left( E,x\right) +I_{ee}(E,x)+I_{ph}(E,x)=0
\label{diffeq}
\end{equation}
where $D$ is the diffusion coefficient of the electrons.\cite{Nagaev2} The
boundary conditions are given by Fermi-Dirac distributions with 
$f\left(E,0\right) =\left[ \exp(E/kT) + 1   \right]^{-1}$ for the left reservoir and 
$f\left(E,L\right) =\left[\exp\left( (E-eV) /kT\right)+1\right]^{-1} $ for the right
reservoir. It is assumed that the reservoirs keep the two ends of the wire at
constant electrochemical potential $0$ and $eV$, resp., and at a constant
temperature $T$ (Fig.~\ref{semiclass} bottom left). If inelastic scattering is absent
(non-interacting regime), the solution of Eq.~(\ref{diffeq}) is a linear
combination of the two reservoir distribution functions ($0\leq x\leq L$):

\begin{equation}
f\left( E,x\right) =\frac{L-x}{L}f\left( E,0\right) +\frac{x}{L}f\left(
E,L\right)  \label{superpos}
\end{equation}
which has the shape of a two-step function (Fig.~\ref{semiclass} bottom
right). Inserting this into Eq.~(\ref{Nagaev}) one obtains for the noise:

\begin{equation}
S_{I}=\frac{2}{3}\left[ \frac{4kT}{R}+\frac{eV}{R}\coth \left( \frac{eV}{2kT}
\right) \right]   \label{onethirdthermal}
\end{equation}
This equation is identical to the result obtained with the 
Landauer-B\"{u}ttiker 
formalism and describes the crossover from thermal noise at $V=0$ to
an asymptotic shot noise behaviour $S_{I}=\frac{1}{3}\cdot 2e\left| I\right| 
$ for $eV\gg kT$. As mentioned above, the same reduction factor also results
from a model using sequential tunneling through a series of tunnel barriers.
Although various theories predict a universal $1/3$-noise reduction factor for the
non-interacting regime, no experiment has yet confirmed the $1/3$-slope in
the asymptotic limit $eV\gg kT$.

Another special case arises if $L\gg l_{e-e}$. The electrons can exchange
energy among each other and are therefore in a local thermodynamic
equilibrium. Hence, the occupation probability $f(E,x)$ is described by a
Fermi-Dirac distribution with a {\em local} electron temperature $T_{e}(x)$
at the electrochemical potential $\mu (x)=\frac{x}{L}eV$:

\begin{equation}
f\left( E,x\right) =\frac{1}{e^{\frac{E-\mu (x)}{kT_{e}(x)}}+1}
\label{fdheating}
\end{equation}
The temperature profile $T_{e}(x)$ along the wire can again be calculated from
Eq.~(\ref{diffeq}), which reduces to a heat-flow equation:

\begin{equation}
\frac{\pounds _{0}}{2}\frac{d^{2}T_{e}^{2}}{dx^{2}}=-\left( \frac{V}{L}
\right) ^{2}+\Gamma \left( \frac{k}{e}\right) ^{2}\left(
T_{e}^{5}-T^{5}\right)   \label{heatdiff}
\end{equation}
where $\pounds _{0}=\frac{\pi ^{2}}{3}\left( \frac{k}{e}\right) ^{2}$ is the
Lorenz number and $\Gamma $ is a parameter describing electron-phonon
scattering. Eq.~(\ref{Nagaev}) turns now into $S_{I}=4k\left\langle
T_{e}\right\rangle _{x}/R$. Hence, the excess noise is now solely due to
thermal noise of the hot electrons and $S_{I}$ is determined by the electron
temperature averaged over the whole wire length. For $L\ll l_{e-ph}$ the
electron-phonon term can be neglected and an analytical solution exists for
the temperature profile (inset Fig.~\ref{twocurves}):\cite{deJong3}

\begin{equation}
T_{e}(x)=\sqrt{T^{2}+\frac{x}{L}\left( 1-\frac{x}{L}\right) \frac{V^{2}}{
\pounds_{0} }}  \label{tempprof}
\end{equation}
This leads to:

\begin{equation}
S_{I}=\frac{2kT}{R}\left[ 1+\left( \nu +\frac{1}{\nu }\right) \arctan \nu 
\right]   \label{sqrt34}
\end{equation}
with $\nu =\sqrt{3}eV/2\pi kT$. For $eV\gg kT$ one obtains $S_{I}=\sqrt{3}
/4\cdot 2e\left| I\right| \simeq 0.43\cdot 2e\left| I\right| $.

Fig.~\ref{twocurves} displays the expected noise versus applied voltage for the
non-interacting regime according to Eq.~(\ref{onethirdthermal}) and for the 
interacting electron picture according to Eq.~(\ref{sqrt34}). 
Both curves start at $V=0$ with thermal noise and
separate into two straight lines with different slopes for $eV\gg kT$. The
figure suggests that at least $eV/kT\gtrsim 10$ is necessary in order to
distinguish the two regimes by the measured asymptotic slopes. An experiment
under such highly non-equilibrium conditions requires special care 
in the treatment of dissipation 
due to the large unavoidable power input. In particular, one has to
consider how energy is removed in the reservoirs attached to the wire.

\subsection{Reservoir heating}

The theory described above assumes ideal boundary conditions for the
electrons at the immediate wire end. The electrons in the
reservoirs are described by a Fermi-Dirac distribution with a constant
electrochemical potential $\mu $\ and a constant bath temperature $T$ 
independent of the current flowing through the wire. 
This assumption is only correct for reservoirs of infinite size with infinite 
electric and heat conductivities.
For real reservoir materials, e.g. Au, Ag, Cu, the actual size and heat
conductance of the
reservoirs will matter. In the following we discuss the different
contributions that can give rise to a temperature increase in the reservoirs
caused by the generated power $V^{2}/R$ which has to dissipate in the
reservoir and substrate. It will turn out that noise is substantially
affected in the non-interacting regime, if the reservoir temperature rises.
The heat flows through a chain of different thermal resistors
connected in series (see Fig.~\ref{heatchain}). We start at the top of the heat
chain where the electronic heat spreads out radially into the whole
reservoirs. We take the radius of the two inner semicircles to be $
r_{1}=l_{e-e}/2$. For the non-interacting regime these semicircles may be
considered as part of the wire itself (the inner white part in 
Fig.~\ref{heatchain}). 
This is justified since the $1/3$-suppression has been shown to hold
independent of the wire geometry, as long as the wire is shorter than $
l_{e-e}$.\cite{Eugene} Since a change in temperature is only defined
over distances larger than $l_{e-e}$, we assume a constant temperature in
this inner region. This is the highest temperature and denoted with $
T_{e,hi} $. Going radially outwards, the power spreads by electronic heat
diffusion in the electron gas which is described by a thermal spreading
resistance $R_{e-diff}$, similar to well known electrical spreading
resistances. The transfer of energy from the electron gas to phonons in the
reservoirs can be neglected up to a radius of order $l_{e-ph}$. For higher
radii the electron-phonon scattering length provides a natural cutoff for
the electronic heat diffusion. We therefore define the largest radius 
$r_{0}$ to be the smaller of either $l_{e-ph}$ or the planar reservoir size
$L_{res}$.
At this distance the electron temperature has dropped to $T_{e,lo}$. In the
heat-chain model, the thermal resistance for the conversion of energy from
electronic to lattice degrees of freedom follows next. First, energy flows
into the phonon system of the reservoir resulting in a difference between 
$T_{e,lo}$ and the reservoir phonon temperature $T_{ph}$. The corresponding
thermal resistance is denoted by $R_{e-ph}$. Then, a thermal-boundary
resistance $R_{K}$ (Kapitza resistance) may give rise to a difference in
phonon temperatures of reservoir $T_{ph}$ and substrate $T_{sub}$. Finally,
the generated heat is transferred into the cryogenic bath, held at the
constant bath temperature $T_{bath}$. This thermal anchor to the bath has
the thermal resistance $R_{s}$. The temperature difference over each
thermal resistor is proportional to the thermal resistance and the power $P$
flowing through it.
The minimization of all thermal resistances
in the complete heat chain is essential to prevent   $T_{e,hi}$ to rise
and thus to prevent the injection of hot electrons into the wire. 
This is in particular
important for the non-interacting regime, since it turns out, that a
temperature rise in this regime results in substantial additional noise in 
the asymptotic limit. 
This can be understood from the asymptotic behaviour of 
Eq.~(\ref{onethirdthermal}) for $eV\gg kT$ which contains a 
temperature dependent
offset in addition to the term linear in $I$:

\begin{equation}
S_{I}=\frac{1}{3}2e\left| I\right| +\frac{8}{3}kT/R  \label{onethirdasym}
\end{equation}
For the temperature $T$, we have to insert $T_{e,hi}$ into 
Eq.~(\ref{onethirdasym}) as the temperature of the injected electrons.
If $T_{e,hi}$ scales linearly with the current $I$, the measured slope will
be larger than $1/3$. It is quite remarkable that in the interacting regime
an increase of $T_{e,hi}$ has only a minor effect for the measured noise. 
As the linear asymptote
for $eV\gg kT$ passes through the origin, the correction to the slope is
only of second order in $kT/eV$.

Next we estimate the increase of the four temperatures in the heat
chain $T_{sub}$, $T_{ph}$, $T_{e,lo}$ and $T_{e,hi}$, when a heat current
flows through the chain. The connection between sample and cryogenic bath
determines the increase of $T_{sub}$. We will see later in the experimental
section that its dependence on the power $P$ is 
phenomenologically best described as:

\begin{equation}
T_{sub}=\left( T_{bath}^{2}+a\cdot P\right) ^{1/2}
\end{equation}
where $a$ describes the thermal coupling of the sample to the cryogenic bath.

A possible difference between $T_{ph}$ and $T_{sub}$ is due to a Kapitza
resistance and can be written as:\cite{Wellstood} 
\begin{equation}
T_{ph}=\left( T_{sub}^{4}+\frac{P}{A\sigma _{K}}\right) ^{1/4}
\end{equation}
$A$ denotes the area of the reservoir and $\sigma _{K}$ is a parameter 
specific for the interface between reservoir and substrate. 
Because of the large size of the reservoirs in this work, this is a small 
effect, but was added here for completeness.

To calculate the difference between electron temperature $T_{e,lo}$ and
phonon temperature $T_{ph}$ in the reservoir, we assume for simplicity that
the electron temperature is constant over the whole reservoir. When we
multiply Eq.~(\ref{heatdiff}) with the electrical conductivity $\sigma $,
the second term on the right-hand $\sigma \left( \frac{k}{e}
\right) ^{2}\Gamma \left( T_{e,lo}^{5}-T_{ph}^{5}\right) $
is the power per volume
dissipated by electron-phonon scattering and can be set
equal to $\frac{V^{2}}{R A t }$. $\sigma $ is now the electrical conductivity 
of the reservoir and $t$ its thickness. We obtain:
\begin{equation}
T_{e,lo}=\left( T_{ph}^{5}+\frac{V^2}{R}\frac{R_{\Box}}{\Gamma A }
\left( \frac{e}{k}\right) ^{2}\right) ^{1/5} \label{eph}
\end{equation}
where we have introduced the sheet resistance of the reservoir $R_{\Box}=
1/\left( \sigma t\right) $. 
The parameter $\Gamma $ is known from noise 
measurements on long diffusive wires
($L\gg l_{e-ph}$)\cite{Wellstood,Roukes,Henny} 
and can be used to determine the electron-phonon
scattering length\cite{Wellstood} $l_{e-ph}=1.31/\sqrt{T^{3}\Gamma }$.

Finally, in order to determine the temperature in the wire $T_{e,hi}$, we
have to calculate the temperature gradient in the reservoir due to radial
electronic heat diffusion from the inner semicircles with radius $r_{1}$
to the outer ones with radius $r_{0}$ (see Fig.~\ref{heatchain}). 
Using cylindrical symmetry the heat flow density is given by

\begin{mathletters}
\begin{equation}
\overrightarrow{j}=-\kappa \overrightarrow{\nabla }T=\frac{P}{2\pi rt}
\overrightarrow{e}_{r}  \label{WF}
\end{equation}
where $r$ is the radius of a semicircle between $r_{1}$ and $r_{0}$, $t$ the
thickness of the reservoir and $\kappa $ the electronic thermal conductivity
derived from the Wiedemann-Franz law $\kappa =\pounds _{0}T\sigma $,
the latter has been shown to be valid in small wires.\cite{Steinbach,Henny}
Integrating over the temperature gradient $\overrightarrow{\nabla }T$\ with
the boundary condition $T\left( r_{0}\right) =T_{e,lo}$, 
yields for $T_{e,hi}=T\left( r_{1}\right) $:

\begin{equation}
T_{e,hi}^{2}=T_{e,lo}^{2}+b^{2}V^{2}  \label{ediff}
\end{equation}
with
\begin{equation}
b= \sqrt{\frac{1}{\pi \pounds _{0}}\frac{R_{\Box }}{R}\ln \frac{
r_{0}}{r_{1}}} \label{bdef}
\end{equation}
For large applied voltages, 
the second term on the right-hand
side of Eq.~(\ref{ediff}) dominates and a linear dependence
of the electron temperature with respect to the applied voltage is obtained:
\begin{equation}
T_{e,hi}=b\cdot V  \label{Tehi}
\end{equation}
When inserting Eq.~(\ref{Tehi}) into Eq.~(\ref{onethirdasym}), the increase
in noise $\Delta S_{I}$ can be calculated and one obtains for the additional
slope:

\begin{equation}
\frac{\Delta S_{I}}{2eI}=\frac{4}{3}\frac{k}{e}\cdot b=
\frac{4}{3}\sqrt{\frac{3}{\pi ^{3}}\frac{R_{\Box}}{R}\ln \frac{r_{0}}{r_{1}}}  
\label{slopeincrease}
\end{equation}

Hence in the independent-electron regime the measured slope is always
larger than $1/3$! The increase in slope is determined by the ratio 
$R_{\Box}/R$ and the geometrical parameters $r_{0}$ and $r_{1}$. 
The electrical
parameters $R_{\Box }$ and $R$ are known accurately. 
For the radii natural cutoffs have been introduced: $l_{e-e}/2$ for
$r_1$ and the smaller of either $l_{e-ph}$ or the reservoir size
$L_{res}$ for $r_0$.
Though the assumed values for $r_0$ and $r_1$ are correct on physical 
grounds, a more rigorous theory may give a slightly different prefactor.
Since $r_0$ and $r_1$ enter 
Eq.~(\ref{slopeincrease}) only logarithmically, corrections are small. 
Both $l_{e-e}(T)$ and $l_{e-ph}(T)$ display a power-law 
dependence on temperature $T$
effectively resulting in the cutoff term $ln \left(r_0/r_1 \right)$ to be
temperature dependent as well, albeit weakly, only proportional to $ln(T)$.
This weaker temperature dependence will be neglected in the following. For 
$r_0$ and $r_1$ values typical for the experiment  will be used. 

In the following we compare the magnitude of the temperature increase
caused by electronic heat diffusion using Eq.~(\ref{ediff}) and 
electron-phonon scattering using Eq.~(\ref{eph}).
In Fig.~\ref{crossover} the relative increase $\Delta T/T$ is plotted as a 
function of bath temperature $T$ for fixed $eV/kT=20$, which is a typical
value used to distinguish between the interacting and non-interacting regime. 
Within the above mentioned assumption, the contribution from
electronic heat diffusion is independent of $T$, the two plotted values
(dashed lines)
correspond to a ratio of $R/R_{\Box }=250$ and $R/R_{\Box }=1000$
with $r_0/r_1 = 100$. In
contrast, the electron-phonon coupling strongly depends on $T$. Its 
thermal resistance increases with decreasing temperature, since
the electron-phonon scattering rate is proportional to $T^3$. 
This results in a drastic increase of $\Delta T/T$ at low temperatures
in Fig.~\ref{crossover} (solid curves correspond to different lateral
reservoir sizes as denoted).
Due to this sharp rise the study of non-equilibrium effects at very low 
temperatures becomes increasingly difficult.\cite{Roukes} 
The large temperature increase due to the vanishing coupling of the electrons
to phonons at low temperatures can only be compensated
by enlarging the reservoir volume. Note that 
both contributions depend on the reservoir thickness, which is
included in the reservoir sheet resistance $R_{\Box }$.

Up to now, as a first approximation, we have treated electronic heat diffusion
and electron-phonon scattering independently. This is
certainly not fully correct. The electron temperature, which is relevant for
the electron-phonon scattering, is not constant over the reservoir as
previously assumed.
To determine the temperature profile self-consistently,
we can combine the electronic heat
diffusion and the electron-phonon scattering term in one equation, which has
a similar form as Eq.~(\ref{heatdiff}), but now in cylindrical coordinates.
We assume that the voltage drop across the reservoirs is negligible,
so that the heat-generating term can be omitted: 
\end{mathletters}
\begin{mathletters}
\begin{equation}
\frac{\pounds _{0}}{2}\left[ \frac{d^{2}T_{e}^{2}}{dr^{2}}+\frac{T_{e}}{r}
\frac{dT_{e}}{dr}\right] =\Gamma \left( \frac{k}{e}\right) ^{2}\left(
T_{e}^{5}-T_{ph}^{5}\right)   \label{heatdiffres}
\end{equation}
The power enters the system at a semicircle of radius $r_{1}$ defining
the first boundary condition. According to Eq.~(\ref{WF}) 
it is given by:
\begin{equation}
\frac{\pounds _{0}}{2}\frac{d}{dr}T^{2}\left( r_{1}\right) 
=\frac{P}{2\pi r_{1}}R_{\Box }
\label{bound1}
\end{equation}
We now assume the reservoir to be terminated by a semicircle of radius 
$r_{0}$. The heat flow at the end of the reservoir must vanish and the
second boundary condition reads: 
\begin{equation}
\frac{\pounds _{0}}{2}\frac{d}{dr}T^{2}\left( r_{0}\right) =0
\label{bound2}
\end{equation}
\end{mathletters}
The differential equation (\ref{heatdiffres}) together with the boundary
conditions (\ref{bound1}) and (\ref{bound2}) cannot be solved analytically. 
To obtain
quantitative estimates for $T_{e,hi}$, we have performed a simulation using
the method of finite elements. 
We have varied the power $P$, the
electron-phonon scattering parameter $\Gamma $ and the reservoir outer
and inner radii $r_{0}$ and $r_{1}$. The main results are as follows:
The electron temperature decays approximately exponentially from $T_{e,hi}$
at the inner radius $r_{1}$ to a base temperature $T_{e,lo}$ at $r_{0}$. 
The decay length,
over which $T_{e}-T_{ph}$ is reduced by a factor $e$, is about $l_{e-ph}/4$,
where $l_{e-ph}$ is the electron-phonon scattering length at $T_{e,lo}
\simeq T_{ph}$. The
resulting temperature profile of two simulations with different $\Gamma $ is
plotted in Fig.~\ref{simulation}. The inset shows the difference $T_{e}-T_{ph}$
on a logarithmic scale. 
The two straight slopes indicate the exponential decay
of the temperature $T_{e}$ to $T_{ph}$. The decay length depends only
slightly on the power $P$.
If $r_{0}\gtrsim 2\cdot l_{e-ph}$ no significant raise of $T_{e,lo}$ with
respect to $T_{ph}$ is found and $T_{e,hi}$ depends only on the
incoming power and the reservoir sheet resistance. 
This corresponds to the limit described above, where the electron-phonon
contribution is small compared to the one from electronic heat diffusion. 
It can be used as a design criteria for reservoirs appropriate in minimizing
dissipative reservoir heating.
The reservoir size, which
is required for this is plotted as a function of $T_{ph}$ in the inset of
Fig.~\ref{heatprediction}. 
The simulation also shows that
the functional behaviour of $T_{e,hi}$ with applied voltage $V$ can be 
described like in Eq.~(\ref{ediff}). 
The relation $b\propto \sqrt{R_{\Box }/R}$ is still valid consistent
with Eq.~(\ref{bdef}) and the proportionality factor corresponds to a ratio of 
about $r_{0}/r_{1}=100$, which is very reasonable. Such a ratio
would also follow from our simple analytical model, 
when the electron-phonon scattering length 
is inserted for  $r_{0}$
and  the electron-electron scattering length for $r_{1}$ taken at subkelvin
temperatures. This discussion shows that large reservoirs
are needed to minimize the increase in reservoir temperature.
In particular, if $L_{res}\gg  l_{e-ph}$ is followed in the
design of the reservoirs, the main contribution for the relative
temperature rise $\Delta T/T$ is caused by electronic heat diffusion,
which is displayed in Fig.~\ref{heatprediction}  
as a function of the applied voltage for three different 
ratios of $R/R_{\Box }$. 
As can be seen, the temperature increase can be substantial.

\section{Experiment}

\subsection{Design}

In the experiments described below we explore the $1/3$-shot noise
suppression in the non-interacting regime and study the influence of
different reservoir configurations. In view of the important role of the
reservoirs discussed above a careful design of the experiment is crucial. 
The non-interacting regime requires $L\ll l_{e-e}$.
For an estimate of $l_{e-e}$ we use Altshuler's formula valid for a
one-dimensional wire:

\begin{equation}
l_{e-e}=\left[ \left( \frac{\sqrt{2}}{k_{B}}\right) \left( \frac{\hbar }{e}
\right) ^{2}\frac{D\cdot w}{T\cdot R_{\Box }^w}\right] ^{1/3}
\label{Altshuler}
\end{equation}
where $w$ is the width and $R_{\Box }^w$ the sheet resistance of the 
wire.\cite{Altshuler} 
For a typical Au wire with a thickness of $15$~nm,
diffusion coefficient $D=120$~cm$^{2}$/s, width $w=100$~nm and $R_{\Box
}^w=2.3$~$\Omega $, we find a scattering length $l_{e-e}=4.2~\mu $m at $0.3$~K.
Using standard e-beam lithography, a wire with a length of $1$ $\mu $m
connected to two reservoirs is feasible. Shorter wires are
difficult to fabricate because of proximity-effect from exposing the large 
areas of the two reservoirs.

As mentioned above, in order to distinguish the $1/3$ from the 
$\sqrt{3}/4$-regimes a ratio of at least $eV/kT\gtrsim 10$ is necessary. 
A low base temperature is required, since
otherwise the applied voltage becomes too high and electron-phonon scattering
in the wire is unavoidable. 
To get an estimate of the influence of electron-phonon 
scattering on noise we have to compare the wire length 
with the electron-phonon scattering length at temperature $eV/k$. 
We find that a deviation in noise of about 1$\%$ would result 
if $L\simeq 4\cdot l_{e-ph}$. For a $1$~$\mu $m long wire with 
$\Gamma =5\cdot 10^{9}$~m$^{-2}$K$^{-3}$ this relates to 
a maximum voltage, which
corresponds to $17.6$~K. For a ratio of $eV/kT=40$
(the largest ratio used in the experiment), the 
bath temperature shall be lower than $440$~mK.
As explained above the reservoir heating strongly depends on the ratio 
$R_{\Box }/R$, which ought to be as small as possible to avoid heating. 
In our experiment we will vary this ratio. As we
have fixed the length of the wire, its width and thickness should be 
small to achieve a high wire resistance. On the other hand, the reservoirs have
to be as thick as possible and made of a highly-conductive metal to reduce
$R_{\Box }$.

The size of the reservoirs has to be chosen according
to the electron-phonon scattering length 
$l_{e-ph}$ in the reservoir. Its radius $r$ should be about twice $l_{e-ph}$
to avoid a significant difference between electron and phonon temperature,
see inset of Fig.~\ref{heatprediction}.
With $\Gamma =5\cdot 10^{9}$~m$^{-2}$K$^{-3}$ we obtain 
$l_{e-ph}=1.31/\sqrt{T_{e}^{3}\Gamma }=110~\mu $m at $0.3$~K, 
which means that two rectangles with $200~\mu $m~$\times ~400~\mu $m 
on each side are sufficient.\cite{Henny} 
Note, that for $50$~mK this length even exceeds $1.5$~mm.

An estimate has also to be made for the temperature increase due to a Kapitza
resistance. As a worst case estimate for $\sigma _{K}$ we use 
$100$~W/m$^2$K$^4$.
With a heating power of $50$~nW we expect a temperature increase of only 
$25$~mK, which is small compared to the applied voltage $eV/k=12$~K.

In our experiment, a possible increase of the substrate temperature $T_{sub}$ is 
taken into account, since we can measure $T_{sub}$ directly with noise 
thermometry using an additional monitor wire on the same substrate.

\subsection{Sample Fabrication}

The samples were produced with standard e-beam lithography. A $600$~nm thick
PMMA-resist was spun on an oxidized Si(100)-wafer and structured with a JEOL
JSM-IC 848 at an acceleration voltage of $35$~kV. The pattern consisted of a
line (line dose $\sim 1.8$~nC/cm) and of two areas on each side of the line. To
correct for the proximity-effect the area dose was increased in steps from the wire
ends ($\sim 200~\mu $C/cm$^{2}$) to the outer part of the reservoirs ($400$~$
\mu $C/cm$^{2}$). The small structures were written with a probe current of $
40$~pA and the large pads with $16$~nA. In order to enable a second
lithography step 8 alignment marks were written. This structure was repeated
up to $40$ times on the same substrate. The resist was developed in $
MiBK:IPA=1:3$ during $45$~s. Metal evaporation was performed with the
two-angle evaporation technique.\cite{Henny} 
First a $15$~nm Au-layer was evaporated
under normal incidence. Then for the reservoirs a second $200$~nm Au-layer was
evaporated at a tilt angle of $30^{\circ}$ without breaking the 
vacuum. 
This ensures a good contact between the wire and the reservoir.
Even larger reservoirs were produced in a second lithographic step
in which $1$~$\mu$m thick Cu layers were aligned over the previous reservoirs.
Relevant parameters of the three samples are summarized in 
Table~\ref{Parameters}.

\subsection{Noise Measurement Setup}

The lower inset of Fig.~\ref{setup} shows the noise measurement set-up. The
sample with resistance $R$ is biased by a current provided by the constant
voltage source connected to large series resistors $R_{s}\gg R$. 
The voltage over the 
sample is then amplified with a gain of $1000$ by two independent low-noise
preamplifiers (EG\&G 5184) operated at
room temperature. The noise spectrum is obtained by a
cross-correlation of the two amplifier signals using a spectrum analyzer (HP
89410A). 
This correlation scheme effectively removes voltage-offset noise from
the preamplifiers.\cite{Glattli}
For every data point the signal is averaged over a frequency
bandwidth of $70$~kHz at
a typical center frequency of $300$~kHz (at this frequency
$1/f$-noise is absent).
With a measuring time of $60$~s a
sensitivity of $10^{-22}$~V$^{2}$s is achieved. As we measure the voltage
fluctuations $S_{V}=S_{I}\cdot R^{2}$, the signal $S_{V}=\frac{1}{3}2eV\cdot
R$ is proportional to $R$. We aimed at a precision of $1\%$ at a ratio $
eV/kT=40$, which gives us a lower limit for the sample resistance of $R=90$~$
\Omega $. Within the geometrical requirements the typical resistance is however
in the range of $10-20$~$\Omega $. To increase the sample resistance and
with it the precision, we use a series of many identical wires, all attached to
individual reservoirs. The
resistance of each wire was first measured at room temperature to obtain the
scattering $\Delta R$ around the average resistance $\overline{R}$.

For an absolute noise measurement, a calibration of the complete
setup is unavoidable. The measured noise signal is affected
by shunt capacitances from the leads in the cryostat which partially diminish
the dynamical signal.
We calibrate the measured excess noise against the thermal noise of the
same sample measured within the same frequency bandwidth.
This is done for every sample separately, since the resistance varies from
sample to sample.
A typical calibration is shown in Fig.~\ref{setup}. 
The thermal noise of the sample varies linearly with  temperature $T$ according
to $S_U=4kTR$  with an offset, which arises from current noise of the
preamplifiers. Since the resistance $R$ is known from an independent
DC-measurement, the slope and offset of the line in Fig.~\ref{setup}
provides us with the absolute calibration.

As mentioned above, the substrate heating is determined from the thermal noise
of an unbiased monitor wire on the same substrate. 
A typical measurement is displayed in the upper inset of Fig.~\ref{setup}. 
The dependence of the data could best be accounted for
by the phenomenological relation
$T_{sub}=\left( T^{2}+a\cdot P\right)^{1/2} $.
It yields as fit parameter $a=1.31\cdot 10^{5}$~K$^{2}$/W, 
which is specific for the cryostat.

\section{Results and Discussion}

We now discuss the experimental results for three different samples which 
mainly differ in the heat conductance of their reservoirs.
In Fig.~\ref{measurement} the measured shot noise of the samples (AI,
B, AII) is plotted. The solid lines are calculated assuming non-interacting
electrons (lower curve, slope $1/3$) and interacting electrons (upper curve,
slope $\sqrt{3}/4$). Two corrections are included in these theoretical
lines: the increased substrate temperature using the parameter $a$ and the
relative scattering of the wire resistances around its average $\Delta R/
\overline{R}$, which has however only a small influence of around 1\%. The
relevant sample parameters are summarized in Table~\ref{Parameters}.

Sample AI consists of 28 wires with an average resistance of 
$\overline{R}=11.8~\Omega $
and $200~$nm thick Au reservoirs resulting in a reservoir 
sheet resistance of  $R_{\Box}=42~$m$\Omega $. 
In Fig.~\ref{measurement}a the measured noise of this sample as a function of 
current is shown. Within the accuracy of the experiment,
the data points lie on the $\sqrt{3}/4$-curve and one may on first sight
infer that the length of the wire (910~nm) is much longer than the 
electron-electron scattering length in contradiction to Eq.~(\ref{Altshuler}). 
This conclusion is, however, only valid
 if reservoir heating is completely absent.

For sample B the same wire length and reservoir thickness are used. 
Since the wires of this sample are narrower, their resistances $R$ are higher, 
so that we expect to have less heating as compared to sample AI,
since $R_{\Box}/\overline{R}$ is reduced.
As is evident from  Fig.~\ref{measurement}b the measured
noise is indeed much lower lying closer to the $1/3$-curve than 
to the $\sqrt{3}/4$-curve. For the highest
applied voltage we have $eV/kT\simeq 35$ in both cases.

In order to increase $R/R_{\Box }$ even further, we have
fabricated thicker reservoirs with a much lower sheet resistance. 
Sample AII has initially been the same as sample AI, but in a second
lithography step a $1~\mu $m thick Cu layer has been evaporated onto the
reservoirs in addition to a thin Au layer
preventing oxidation of the Cu reservoir (see Fig.~\ref{sempic}). 
This reduces the reservoir sheet resistance
considerably to $2.8~$m$\Omega $.
 During the second processing of the sample,
several wires were lost and only 8 of them remained for the measurement of
sample AII. 
Because of the reduced total resistance $R_{tot}$, the measured noise
voltage is lower, thus increasing the scatter in the data points, 
but a clear reduction
of the slope is visible when comparing 
with the measurement of sample AI
(see Fig.~\ref{measurement}a and Fig.~\ref{measurement}c, respectively).
The data points are now consistent with the $1/3$-prediction.

Since an asymptotic slope of  $1/3$ is the prediction for the non-interacting
electron regime, sample AII has to be in this regime and therefore also
AI (same wires), even though the latter displays a significantly increased
noise indistinguishable from an asymptotic $\sqrt{3}/4$-slope. 
Since the wires used for sample B are made from the same material with a similar
length, sample B must be in the independent regime as well.
All three samples are in the non-interacting regime according to the
theoretical estimate given above. However, only  for sample AII with 
the highest conducting
reservoirs the measured noise corresponds to the prediction for this
regime. For the other two samples additional noise is detected, which 
increases as $R/R_{\Box }$ becomes smaller.

We explain this increase of noise with electron heat diffusion in the 
reservoirs.
Since we have estimated $R_{e-diff}$ to be the dominant thermal resistance 
for all reservoirs in this work, we expect from our model, that the temperature
of the electrons injected into the wires to vary as $T_{e,hi}=\left(
T_{sub}^{2}+b^{2}\cdot V^{2}\right) ^{1/2}$ according to Eq.~(\ref{ediff}).
Inserting this voltage-dependent temperature into Eq.~(\ref{onethirdthermal})
we can treat $b$ as a fit parameter which describes the magnitude of
the heating. If our heating model is valid, $b\propto\sqrt{R_{\Box }/R}$. 
In Fig.~\ref{allinone} the fitted values of $b$ are plotted
as a function of $\sqrt{R_{\Box }/R}$ for the three samples. Within the
error bars it is consistent with the proportionality to  $\sqrt{R_{\Box }/R}$ 
as we have proposed
it with our heating model. The plotted line is a least square fit with the
assumption that for $\sqrt{R_{\Box }/R}=0$ (i.e. ideal reservoirs) 
no heating is present. The values of $b$ are higher by a factor of $1.8$
than expected from our model.
A higher thermal resistance between electron and phonon temperature 
$R_{e-ph}$ would scale with $\sqrt{R_{\Box }/R}$ as well.
Such a contribution can however be ruled out. 
Although the relevant parameter $\Gamma =5\cdot 10^{9}$~m$^{-2}$K$^{-3}$, 
which was obtained in a $20$~nm thick Au film, could be smaller in the reservoir
due to a larger diffusion coefficient, such an increase would be negligible.
A contribution from a Kapitza resistance would be independent
of $R_{\Box }/R$.
A calculation using $\sigma_K=100$~W/m$^2$K$^4$ would explain 
in maximum an increase of $22$~mK corresponding to a change in $b$ of about 
$23$~K/V, which again is negligible. 

In view of the current debate of a possibly enhanced electron-electron 
interaction, it is important to identify whether the additional 
shot noise originates from heating or from electron-electron scattering.
A possible contribution to the noise arising from electron-electron scattering is 
however independent of $R_{\Box }/R$ and would thus shift the values of $b$ 
by a constant offset.
From Fig.~\ref{allinone} we can estimate such a contribution in our data 
to be less than $100$~K/V corresponding to an increase of 
$0.01\cdot 2e\left| I\right|$ in the asymptotic limit (see below). 

The nearly linear dependence of $b$ with $\sqrt{R_{\Box}/R}$ proves
that the major part of the additional noise
in our experiment can solely be explained by thermal 
heating due to a temperature
gradient in the reservoir and that the wires are indeed in the non-interacting
regime. 

Our measurements support experimental results by Schoelkopf {\it et al.} who
compared measured differential noise on short diffusive wires
with the interacting and non-interacting
theories and found good agreement only for the non-interacting regime, hence
the $1/3$-theory. These experiments were performed at lower voltages
where heating effects are less 
important (Fig.~\ref{heatprediction}). 
However, the absolute slope of $S_I$ in the
asymptotic limit could not be extracted in that work. 
An absolute value has been reported by Steinbach {\it et al}. 
The measured slope was however found  to be
significantly larger than $1/3$.
They explained the increase of noise partly by heating and partly by residual
electron-electron interactions and proposed to use the shot noise measurement
for an independent measurement of the electron-electron interaction in thin
metal films.
The uncertainty on how large the electron-electron 
scattering really is, has led
to the experiment by H.~Pothier {\it et al.}\cite{Pothier} who directly measured 
the electron distribution function by tunneling spectroscopy.
Based on those results we have estimated the residual contribution
from electron-electron scattering in our wires.  
The only relevant parameter
is the ratio of the dwell time of an electron in the wire 
$\tau _{D}=L^{2}/D=70$~ps to the scattering parameter $\tau _{0}=1$~ns
from Ref.~\onlinecite{Pothier}.
A numerical simulation is used to calculate the electron-distribution function
$f(E,x)$ in the wire. Inserting this distribution into Eq.~(\ref{Nagaev}), we obtain the
shot noise which is now slightly larger than $1/3\cdot 2e\left| I\right| $
in the asymptotic limit.
This  increase due to electron-electron scattering is however
only of the order of $0.007\cdot 2e\left| I\right| $.
As mentioned above our data displayed in Fig.~\ref{allinone} 
is not in contradiction, since the error bars would allow for an 
offset independent of $R_{\Box }/R$ of the order of $0.01\cdot 2e\left| I\right|$.

\section{Conclusion}

In this work we have shown, that for a metallic diffusive wire a shot 
noise power consistent with the universal value $1/3\cdot 2e\left| I\right| $ 
is experimentally obtained in the asymptotic limit $eV\gg kT$ if the
reservoirs are designed to minimize a temperature rise as current flows
through the wire. This implies that the ratio between wire resistance $R$
and reservoir sheet resistance $R_{\Box }$ should be large, i.e. of the
order of $1000$ to avoid a large temperature gradient 
due to electronic heat diffusion from the wire region into
the reservoirs. The lateral reservoir size is set by the electron-phonon 
scattering length.
To avoid a difference between the electron and phonon temperatures,
the radius of the reservoir should be at least $2l_{e-ph}$.
In a very striking manner, our experiment demonstrate that 
shot-noise reduction factors close to $\sqrt{3}/4$ can be measured
in the asymptotic limit even for wires that {\em must} be in 
the independent-electron regime!
Though we have a hold of the universial $1/3$ noise-suppression
factor for diffusive wires in the non-interacting electron regime,
another lesson can be drawn from the present experiments: In all highly
non-equilibrium electric-transport experiments conducted at
low temperatures one has to include the
complete environment up to macroscopically large
distances. In this respect experiments differ markedly from the
approach of a theorist, who can separate the wire from
the environment by imposing ideal boundary conditions.
However, ideal boundaries (reservoirs) are non-trivial in real experiments! 

\section{Acknowledgements}

We would like to thank H. Pothier for providing us the computer program to
calculate the electron distribution function in the out-of-equilibrium
regime. Fruitful discussions with M. B\"uttiker, D. Loss, and E.V. Sukhorukov
are gratefully acknowledged. This work is supported by a grant from the
Swiss National Science Foundation.

\begin{figure}
\caption{The electron distribution function of a wire connected to two large
reservoirs at its ends is shown for the case of an applied voltage $V$. 
In the reservoirs and at the wire ends 
the distribution function is a Fermi-Dirac distribution at the chemical potential
$0$ and $eV$ (bottom left). Within the wire it is a two-step function 
if no inelastic scattering is present, $L\ll l_{e-e}$ (solid line) or it is a
Fermi-Dirac distribution with an effective electron temperature $kT_e$ being 
of the order $eV$ if $L\gg l_{e-e}$ (dashed line).} 
\label{semiclass}
\end{figure}

\begin{figure}
\caption{Calculated noise power for the non-interacting regime $L\ll l_{e-e}$, 
(lower curve) and for the interacting regime where $L\gg l_{e-e}$ (upper curve). 
To distinguish between the two regimes in the asymptotic limit a ratio of 
at least $eV/kT\simeq 10$ is required.
The inset shows the temperature profile in the interacting regime along the 
wire for $eV/kT=20$. } 
\label{twocurves}
\end{figure}

\begin{figure}
\caption{The power $V^{2}/R$ produced in the wire has to be dissipated in the 
reservoirs and in the substrate. 
For that it has to pass a series of thermal resistors. 
First, it is distributed in the reservoir by diffusion. 
Then, the heat is transferred by electron-phonon scattering into the phonon 
system of the reservoir from where it flows into the substrate and finally 
into the cryogenic bath kept at the constant temperature $T_{bath}$. 
Over every thermal resistor a temperature drop proportional to the 
resistance and power is induced.} 
\label{heatchain}
\end{figure}

\begin{figure}
\caption{The relative temperature difference $\Delta T/T$ is plotted for 
$eV/kT=20$ as a function of $T$ for various types of reservoirs. 
The relative increase due to $R_{e-diff}$ (dashed lines) strongly depends 
on the 
ratio wire resistance $R$ to reservoir sheet resistance $R_{\Box }$. 
The contribution from electron-phonon scattering (solid lines) 
is strongly temperature dependent and 
increases with decreasing temperature. 
Its magnitude depends mainly on the reservoir's lateral size 
(denoted next to the curve), the electron-phonon scattering parameter 
(here $\Gamma =5\cdot 10^9$~K$^{-3}$m$^{-2}$) and the ratio 
$R/R_{\Box }$ (here 250).} 
\label{crossover}
\end{figure}

\begin{figure}
\caption{Resulting temperature profile in the reservoir obtained from a 
computer simulation using the method of finite elements. 
For an incoming power of $200$~nW and a reservoir sheet resistance 
$R_{\Box }=42$~m$\Omega$, the electron temperature $T_{e,hi}$ rises from 
$0.3$~K to $0.8$~K. 
The curves are calculated with different electron-phonon scattering parameters:
$\Gamma =5\cdot 10^9$~K$^{-3}$m$^{-2}$ for the dashed line ($l_{e-ph}=110$~$
\mu $m) and $\Gamma =1\cdot 10^9$~K$^{-3}$m$^{-2}$ for the solid line
($l_{e-ph}=250$~$\mu $m). 
The inset shows the logarithmic behaviour of the same graphs but after 
subtracting the phonon temperature of $0.3$~K from the electron temperature.} 
\label{simulation}
\end{figure}

\begin{figure}
\caption{Calculated temperature increase $\Delta T/T$ due to electronic heat 
diffusion as a function of applied voltage. 
A linear variation follows if $eV/kT\gg \sqrt{R/R_{\Box }}$.
The inset shows the lateral reservoir size necessary to prevent a temperature 
increase due to electron-phonon scattering. 
It is given by 4$l_{e-ph}=5.24/\sqrt{T^{3}\Gamma }$ with 
$\Gamma =5\cdot 10^9$~K$^{-3}$m$^{-2}$.} 
\label{heatprediction}
\end{figure}

\begin{figure}
\caption{Thermal noise of sample AI used for the calibration of the noise
measurement set-up sketched in the lower inset. 
The upper inset shows the substrate temperature measured on an additional
unbiased monitor wire as a current flows through 
the sample.} 
\label{setup}
\end{figure}

\begin{figure}
\caption{Shot noise measurements for three different samples with different
ratio $R/R_{\Box }$ at a bath temperature $T_{bath}=0.3$~K. 
The upper line corresponds to the prediction of $L\gg l_{e-e}$ (asymptotic 
slope $\sqrt{3}/4$), the lower one to $L\ll l_{e-e}$ (slope $1/3$). 
The measured noise is significantly increased due to reservoir heating 
depending on $R/R_{\Box }$.} 
\label{measurement}
\end{figure}

\begin{figure}
\caption{SEM micrograph of sample AII. The Au wire is terminated by $200$~nm
thick Au reservoirs. In an overlaid second lithography step an additional layer 
of $1$~$\mu$m Cu is
evaporated to increase the reservoir thermal conductance.} 
\label{sempic}
\end{figure}

\begin{figure}
\caption{The parameter $b$, which describes the enhancement of the measured 
noise by heating, is extracted from the data of Fig.~\ref{measurement}. 
It is proportional to $\sqrt{R_{\Box }/R}$. 
The origin of the graph corresponds to a slope of $1/3$ expected for ideal 
reservoirs $R_{\Box }=0$.} 
\label{allinone}
\end{figure}

\begin{table}[tbp] \centering
\begin{tabular}{|c||c|c|c|c|c|c|c|c|}
\hline
& $R_{tot}\left[ \Omega \right] $ & $\#$ & $\overline{R}\left[ \Omega \right]
$ & $L\left[ \text{nm}\right] $ & $w\left[ \text{nm}\right] $ 
& $\text{Reservoir}$ & $R_{\Box }\left[ 
\text{m}\Omega \right] $ & $\left( R_{\Box }/\overline{R}\right) ^{1/2}$
\\ \hline\hline
$\text{AI}$ & $329$ & $28$ & $11.8$ & $910$ & $160$ & $200~\text{nm Au}$ 
& $42$ & $
0.060$ \\ \hline
$\text{B}$ & $129$ & $6$ & $21.5$ & $940$ & $100$ & $200~\text{nm Au}$ 
& $42$ & $
0.044$ \\ \hline
$\text{AII}$ & $74.6$ & $8$ & $9.3$ & $910$ & $170$ & 
$200~\text{nm Au + }1~\mu\text{m Cu}$ & $2.8$ & $0.018$ \\ \hline
\end{tabular}
\caption{Sample parameters at $0.3$~K.\label{Parameters}}
\end{table}


\begin{thebibliography}{99}
\bibitem{deJong}  For a recent review, see M.~J.~M. de Jong and
C.~W.~J. Beenakker, {\it Shot noise in mesoscopic systems }in {\it 
Mesoscopic Electron Transport}, edited by L.~P. Kouwenhoven, G.~Sch\"{o}n
and L.~L. Sohn, NATO ASI Series~E, Kluwer Academic Publishing, Dordrecht
(1996) and the references therein.

\bibitem{Johnson}  M.~B.~Johnson, Phys. Rev. {\bf 29}, 367 (1927);
H.~Nyquist, Phys. Rev. {\bf 32}, 110 (1928)

\bibitem{Schottky}  W.~Schottky,~Ann.~Phys.~(Leipzig)~{\bf 57}, 541 (1918)

\bibitem{Birk}  H.~Birk, M.~J.~M.~de~Jong and C.~Sch\"{o}nenberger, Phys.
Rev. Lett. {\bf 75}, 1610 (1995)

\bibitem{ReznikovKumar}  M.~Reznikov, M.~Heiblum, H.~Shtrikman and
D.~Mahalu, Phys. Rev. Lett. {\bf 75}, 3340 (1995); A.~Kumar, L.~Saminadayar,
D.~C.~Glattli, Y.~Jin and B.~Etienne, Phys. Rev. Lett. {\bf 76}, 2778 (1996)

\bibitem{Steinbach}  A.~Steinbach and J.~M.~Martinis, M.~H.~Devoret, Phys.
Rev. Lett. {\bf 76}, 3806 (1996)

\bibitem{Beenakker}  C.~W.~J.~Beenakker and M.~B\"{u}ttiker, Phys. Rev. B 
{\bf 46}, 1889 (1992)

\bibitem{Nagaev1}  K.~E.~Nagaev, Phys. Lett. A {\bf 169}, 103 (1992)

\bibitem{deJong2}  M.~J.~M.~de~Jong and C.~W.~J.~Beenakker, Phys. Rev. B 
{\bf 51}, 16867 (1995)

\bibitem{Landauer}  R.~Landauer, Physica B {\bf 227}, 156 (1996)

\bibitem{Gonzalez}  T.~Gonz\'{a}lez, C.~Gonz\'{a}lez, J.~Mateos and
D.~Pardo, L.~Reggiani, O.~M.~Bulashenko and J.~M.~Rubi, Phys.~Rev.~Lett. 
{\bf 80}, 2901 (1998)

\bibitem{Eugene}  E.~V.~Sukhorukov and D.~Loss, Phys.~Rev.~Lett. {\bf 80},
4959 (1998)

\bibitem{Liefrink}  F.~Liefrink and J.~I.~Dijkhuis, M.~J.~M.~de~Jong,
L.~W.~Molenkamp and H.~van Houten, Phys. Rev. B {\bf 49}, 14066 (1994)

\bibitem{Altshuler}  B.~L.~Altshuler, A.~G.~Aronov and D.~E.~Khmelnitskii,
J. Phys. C {\bf 15}, 7367 (1982)

\bibitem{Schoelkopf}  R.~J.~Schoelkopf, P.~J.~Burke, A.~A.~Kozhevnikov,
D.~V.~Prober and M.~S.~Rooks, Phys.~Rev.~Lett. {\bf 78}, 3370 (1997)

\bibitem{Pothier}  H.~Pothier, S.~Gu\'{e}ron, N.~O.~Birge, D.~Est\`{e}ve and
M.~H.~Devoret, Phys. Rev. Lett. {\bf 79}, 3490 (1997)

\bibitem{Kogan}  Sh. Kogan, {\it Electronic noise and fluctuations in solids}, 
Cambridge University Press (1996)

\bibitem{Liu}  R.~C.~Liu, B.~Odom, Y.~Yamamoto and S.~Tarucha, Nature {\bf 
391}, 263-265 (1998)

\bibitem{Nagaev2}  K.~E.~Nagaev, Phys. Rev. B {\bf 52}, 4740 (1995)

\bibitem{deJong3}  M.~J.~M.~de~Jong, ''Shot Noise and Electrical Conduction
in Mesoscopic Systems'', Thesis Leiden University, 1995

\bibitem{Wellstood}  F.~C.~Wellstood, C.~Urbina and J.~Clarke, Phys. Rev. B 
{\bf 49}, 5942 (1994)

\bibitem{Roukes}
M.~L.~Roukes, M.~R.~Freeman, R.~S.~Germain, R.~C.~Richardson and 
M.~B.~Ketchen, 
Phys.~Rev.~Lett.~{\bf 55}, 422  (1985).


\bibitem{Henny}  M.~Henny, H.~Birk, R.~Huber, C.~Strunk, A.~Bachtold, 
M.~Kr\"{u}ger and C.~Sch\"{o}nenberger, Appl. Phys. Lett. {\bf 71}, 773 (1997)

\bibitem{Glattli}  D.~C.~Glattli, P.~Jacques, A.~Kumar, P.~Pari and L.~Saminadayar,
J. Appl. Phys. {\bf 81}, 7350 (1997)



\end{thebibliography}
\end{document}